\documentclass[letter,longauth,traditabstract]{aa}
\usepackage{txfonts}
\usepackage{graphicx}

\begin{document}

\title{The detached dust shells of \object{AQ And}, \object{U Ant}, and \object{TT Cyg}\thanks{\textit{Herschel} is an ESA space observatory with science instruments provided by European-led Principal
Investigator consortia and with important participation from NASA.}}

\author{F.~Kerschbaum\inst{1}
\and D.~Ladjal\inst{2}
\and R.~Ottensamer\inst{1,5}
\and M.A.T.~Groenewegen\inst{3} 
\and M.~Mecina\inst{1}
\and J.A.D.L.~Blommaert\inst{2}
\and B.~Baumann\inst{1}
\and L.~Decin\inst{2,4} 
\and B.~Vandenbussche\inst{2}
\and C.~Waelkens\inst{2}
\and T.~Posch\inst{1}
\and E.~Huygen\inst{2}
\and W.~De Meester\inst{2}
\and S.~Regibo\inst{2}
\and P.~Royer\inst{2}
\and K.~Exter\inst{2}
\and C.~Jean\inst{2}
}

\institute{University of Vienna, Department of Astronomy, T{\"u}rken\-schanz\-stra\ss{}e 17, A-1180 Vienna, Austria\newline \email{franz.kerschbaum@univie.ac.at}
\and
Instituut voor Sterrenkunde, Katholieke Universiteit Leuven, Celestijnenlaan 200 D, B-3001 Leuven, Belgium
\and
Royal Observatory of Belgium, Ringlaan 3, B-1180 Brussels, Belgium 
\and
Sterrenkundig Instituut Anton Pannekoek, University of Amsterdam, Kruislaan 403, NL-1098 Amsterdam, The Netherlands
\and
Institute of Computer Vision and Graphics, TU Graz, Inffeldgasse 16/II, A-8010 Graz, Austria}

\date{31 March 2010/ 12 May 2010}

\abstract
{Detached circumstellar dust shells are detected around three carbon variables using \textit{Herschel}-PACS. Two of them are already known on the basis of their thermal CO emission 
and two are visible as extensions in IRAS imaging data. By model fits to the new data sets, physical sizes, expansion timescales, dust temperatures, and more 
are deduced. A comparison with existing molecular CO material shows a high degree of correlation for TT~Cyg and U~Ant but a few distinct differences
with other observables are also found.}

 
 \maketitle
 
\section{Introduction}

For low- and intermediate-mass stars (of initial mass less than 8~$M_{\sun}$) mass-loss mainly takes place on the 
thermally pulsing asymptotic giant branch (AGB) in slow (typically $5 - 25$~km~s$^{-1}$) winds with large mass-loss rates (up to 10$^{-4}$~$M_{\sun}$~yr$^{-1}$).
Whereas the gaseous component of the mass-loss can be observed from the ground (mainly on the basis of molecular emission lines in the mm- and sub-mm radio range), 
the interesting dusty fraction of the wind is preferably observed with infrared space telescopes (see e.g., the review by Olofsson, \cite{Olofs03} and references therein).
IRAS was the first infrared space observatory to enable the identification of a large number of AGB stars with extended dust shells. Young et al. (\cite{Young93b}) interpret them in terms of an interaction of the expelled material with the ISM.

In the generally accepted picture, mass-loss increases on average during the evolution along the AGB (Habing \cite{Habin96}). Nevertheless, observational evidence implies a more episodic mass-loss evolution (van der Veen \&{} Habing \cite{VeenH88}, 
Olofsson et al. \cite{Olofs88}). Based on IRAS photometry, van der~Veen \&{} Habing, identified a group of 60~$\mu$m excess objects with very low temperature 
dust, Olofsson et al. found -- in the course of a larger CO survey -- stars with obviously detached gas shells. These two indicators of perhaps the same phenomenon
were subsequently used to study this interesting evolutionary phase.

Waters et al. (\cite{Water94}) and
Izumiura et al. (\cite{Izumi95}) reconstructed high resolution images using IRAS data at 60 and 100\,$\mu$m and were able to detect a few 
detached dust shells (e.g., \object{U~Ant}, \object{U~Hya}, or \object{Y~CVn}). Soon after this, with the improved sensitivity of the ISO satellite, Izumiura et al. (\cite{Izumi96}) 
were able to image \object{Y~CVn} at both 90 and 160\,$\mu$m. Nevertheless, imaging of xtended dust emission was hampered notoriously by the low
spatial resolution that space telescopes of diameter 60 or 80\,cm could deliver at long wavelengths ($\ge 60$\,$\mu$m) to probe the cold dust of the 
detached envelopes (compare the Akari and Spitzer results presented in Izumiura \cite{Izumi09} and Geise et al. \cite{Geise10}, respectively). Here the 3.5\,m telescope of the \textit{Herschel} Space Observatory (Pilbratt et al. \cite{Pilbr10}) using PACS (Poglitsch et al. \cite{Pogli10}) is expected to play in a different league.

Over the past two decades, the study of detached shells in the CO radio line emission has turned out to be most fruitful. Single dish surveys (Olofsson \cite{Olofs96}, 
and references therein) have revealed a number of these interesting objects, whereas follow-up interferometric maps have uncovered their detailed structure (Lindqvist et al. \cite{Lindq99},
Olofsson et al. \cite{Olofs00}). The available high spatial resolution has detected remarkably spherical and very thin CO-line emitting shells, which are indicative of quite short 
phases of intense mass-loss probably associated with thermal pulses. In the case of TT~Cyg, the authors speak of a high mass-loss phase, 
lasting only a few hundred years, that occurred 7000 years ago. This basic scenario may be modified by more complex wind--wind interactions (Maercker et al. \cite{Maerc10}).
In  general, the detached molecular shells typically detected are geometrically thin and probably smaller in size than the detached dust shells found by IRAS.
It may be that they have a different origin -- episodic mass-loss on the one hand and wind--interstellar medium (ISM) or wind--wind interaction on the other. Objects such as  U Ant (Izumiura et al. 1997) can exhibit
both effects.

\section{Target selection and data reduction}\label{reduction}

The \textit{Herschel} Mass-loss of Evolved StarS  
guaranteed time key program (MESS\footnote{http://www.univie.ac.at/space/MESS/}, Groenewegen et al., in prep.)
investigates the dust and gas chemistry and the properties of CSEs for a large, representative sample of post-main-sequence objects using both imaging and spectroscopy.

The carbon stars AQ And, U Ant, and TT Cyg observed by MESS and discussed in this Letter display a circular emission ring around the central star (Fig.~\ref{f:obs}) at both 70\,$\mu$m and 160\,$\mu$m. 
The star AQ And is a semi-regular carbon star with $T_\mathrm{eff}=$\,2660\,K (Bergeat et al. \cite{Berge01}). Two black-body spectral energy distribution (SED) fits (Kerschbaum \& Hron \cite{KH96b}, Kerschbaum \cite{Kersch99}) infer that there is a `stellar' component at 2353\,K and a `dust' component at 54\,K. Applying the PL-relations (Groenewegen \&{} Whitelock \cite{Groen96}) to the K band, we derive a distance of 825\,pc.
The stars U Ant is a C-irregular of type Lb with a $T_\mathrm{eff}=$\,2810\,K. The SED fits lead to a `stellar' black-body at 2300\,K and a `dusty' one at 72\,K. We assume a distance of 260\,pc consistent with the Hipparcos parallax.
TT Cyg is a C-semi-regular (SRb) with a $T_\mathrm{eff}=$\,2825\,K. Two-black-body SED fits give peak temperatures at 2478\,K and 51\,K. TT Cyg has an uncertain Hipparcos distance of 510\,pc ($-$150, $+$350\,pc). Using PL-relations as above, we find 436\,pc but for consistency with Olofsson et al.~({\cite{Olofs00}) we use 510\,pc.

The observations were carried out using the Photo Detector Array Camera and Spectrometer (PACS, Poglitsch et al. \cite{Pogli10}) at 70~$\mu$m (blue) and 160~$\mu$m (red) leading to a FWHM resolution of 5.6\arcsec{} and 11.3\arcsec. The scan map observing mode was used to achieve a homogeneous sky coverage of 7.5$\arcmin \times$7.5$\arcmin$ for TT\,Cyg, 14.85$\arcmin \times$14.85$\arcmin$ for U\,Ant and 4.95$\arcmin \times$4.95$\arcmin$ for AQ\,And. In this mode, the telescope is slewed at constant speed (in our case 20$\arcsec$/s) along parallel lines to cover the required area.
Two scan maps were made with an angle of 90~$\deg$ between the two to achieve a homogeneous coverage. 

\begin{figure}
\begin{center}
\resizebox{!}{7.1cm}{\includegraphics{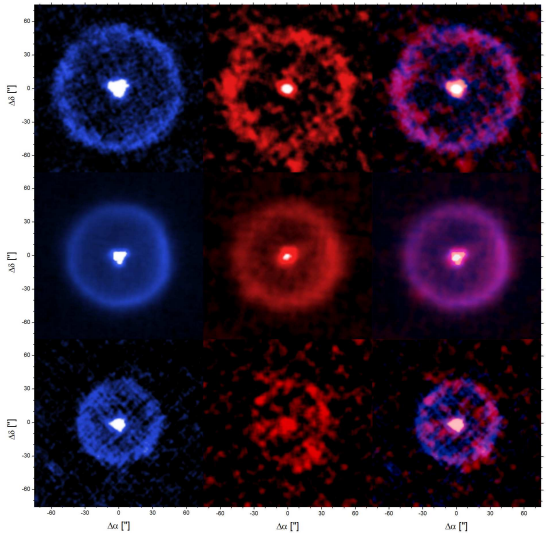}}
\caption{\textit{Herschel}-PACS scan maps (left to right: at 70~$\mu$m, 160~$\mu$m, two colour composite) of AQ~And, U~Ant, and TT~Cyg (top to bottom)}
\label{f:obs}
\vspace{-4mm}
\end{center}
\end{figure}

The \textit{Herschel} interactive processing environment (HIPE\footnote{A joint development by the \textit{Herschel} Science Ground Seg\-ment Consortium, consisting of ESA, the NASA \textit{Herschel} Science Center, and the HIFI, PACS and SPIRE consortia.}) was used for the data reduction, but the nature of our data (bright star surrounded by faint emission) made it necessary to pay special attention to the correlated noise without influencing the photometry of the star. For this purpose, baseline fitting and multiple passes of median filtering were performed. Alternatively, the Madmap module of HIPE was used in addition to custom preprocessing steps that account for detector drifts, to recover very faint extended emission (e.g., the U~Ant far field).
The total flux at PACS wavelength was estimated using aperture photometry and is listed in Table~\ref{flux}.
The aperture radius was selected based on the physical size we see on the intensity profiles (Fig.~\ref{profiles}) and the total flux was corrected following the numbers in the PACS Scan Map release note (PICC-ME-TN-035, 23 Feb 2010) yielding an absolute flux calibration uncertainty of 10-20\%.

\begin{table}
\centering
\caption{\textit{Herschel}-PACS total flux measurements}
\label{flux}
\begin{tabular}{lccc}
\hline
Star   & F$_{70\mu m}$ [Jy]  & F$_{160\mu m}$ [Jy]  & Aperture radius [\arcsec] \\ 
\hline
AQ And & 4.3$\pm$0.7 & 2.3$\pm$0.4 & 65 \\ U Ant & 27.1$\pm$4.1 & 7.4$\pm$1.2 & 60  \\ TT Cyg & 2.9$\pm$0.5 & 0.7$\pm$0.1 & 45  \\ \hline
\end{tabular} 
\vspace{-4mm}
\end{table} 

\section{Analysis} \label{mod}

We assume that the star consists of two spherically symmetric components: an {\em attached} shell reflecting the present day mass-loss and a {\em detached} shell representing the old mass-loss. 
In the optically thin limit the contribution to the SED can be separated into a stellar component F$^{star}_{\nu}$, a present wind-component F$^{wind}_{\nu}$, and detached shell component F$^{shell}_{\nu}$, the total SED then being F$_{\nu}=$F$^{star}_{\nu}$+F$^{wind}_{\nu}$+F$^{shell}_{\nu}$ (see Sch\"oier et al. 2005). More than one detached shell was needed in most cases to fit the data.

\begin{figure}
\begin{center}
\scalebox{0.46}[0.37]{\includegraphics[bb= 54 413 558 705, clip]{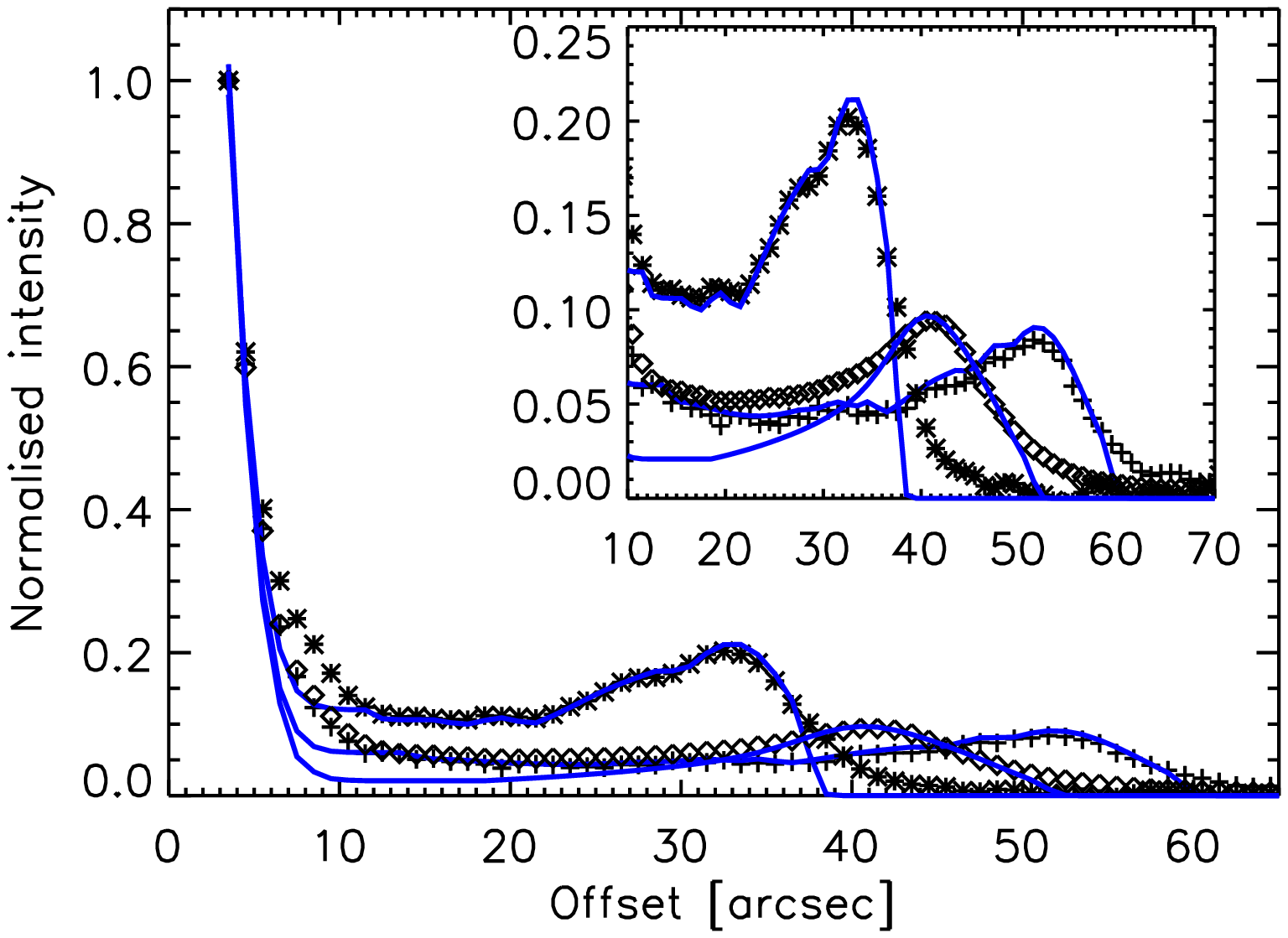}}
\scalebox{0.46}[0.37]{\includegraphics[bb= 54 370 558 694, clip]{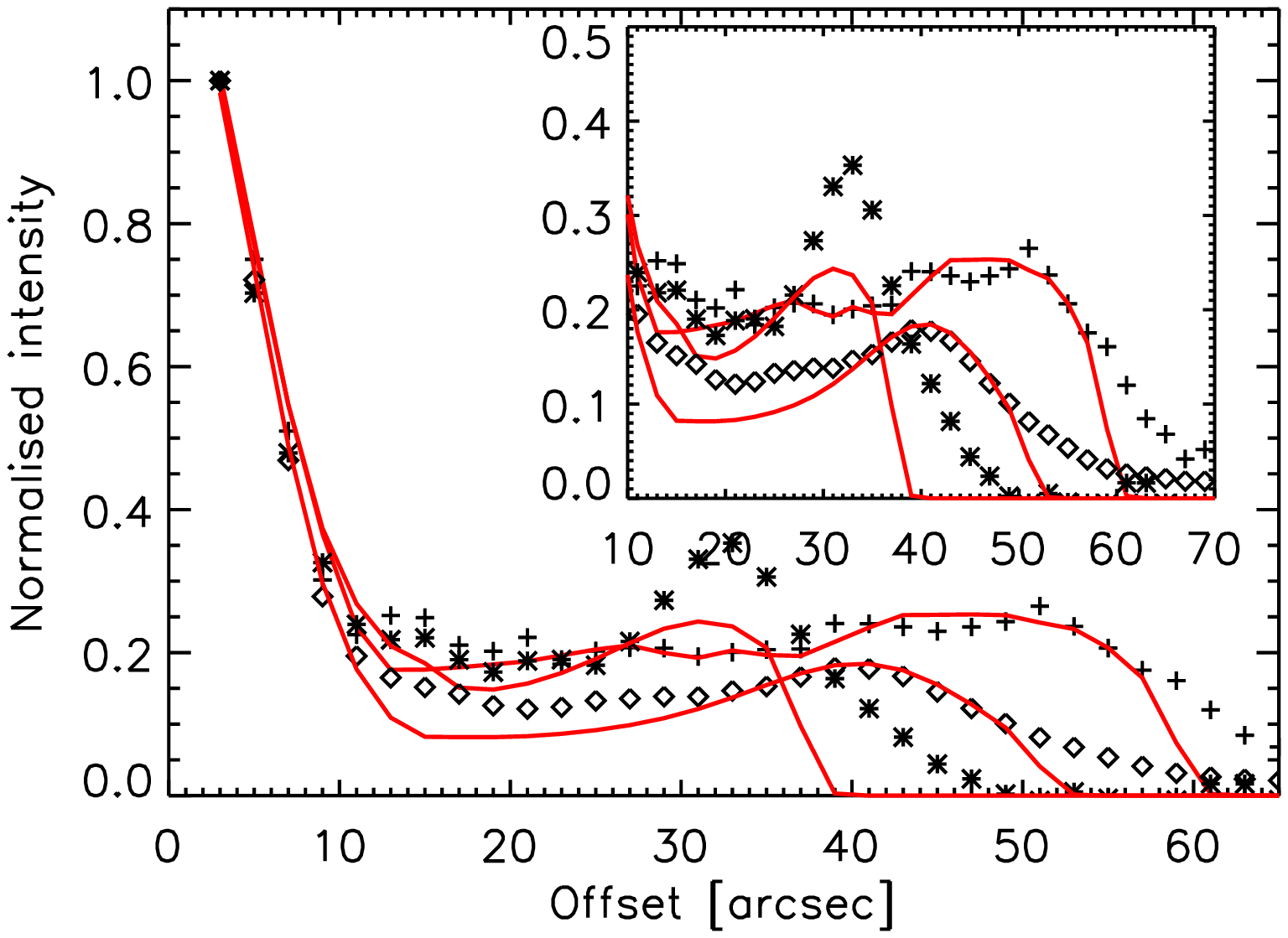}}
\caption{In blue, the 70\,$\mu$m-profiles: crosses indicate AQ And, normalised by 6.14\,mJy/\arcsec{}$^2$, diamonds, U~Ant normalised by 40.27\,mJy/\arcsec{}$^2$, stars, TT~Cyg, normalised by 3.8\,mJy/\arcsec{}$^2$.
In red, the 160\,$\mu$m-filter, normalised by 0.98\,mJy/\arcsec{}$^2$, 5.68\,mJy/\arcsec{}$^2$, and 0.47\,mJy/\arcsec{}$^2$. The lines are the best fits to the data. Inserted is a zoom. 
}
\label{profiles}
\vspace{-4mm}
\end{center}
\end{figure}

Theoretical SEDs and intensity brightness maps were computed using the 1D radiative transfer code DUSTY (Ivezi\'{c} et al. 1999). The continuous distribution of ellipsoids (CDE, see e.g., Min et al. 2003, Hony et al. 2002) was adopted as the grain density distribution. The grain size is such as the volume averaged by a sphere has a radius of 0.05 $\mu$m.
For the chemical composition, we used amorphous carbon with the optical properties of Preibisch et al. (1993) and a density of 1.85 g cm$^{-3}$. For the external radiation source synthetic models of Aringer et al. (2009)  were used.
A grid of attached shells and detached shells was constructed by varying: the temperature of the dust at the inner radius, the optical depth $\mathbf{\tau}_{\lambda}$ at 0.55~$\mu$m, and the outer radius of the shell. A density power index $n$ = 2 was adopted.
We first fitted the synthetic to the observed SED and photometry collected from the literature (Fig.~\ref{SED}) using a least squares minimisation technique to obtain a rough estimate of the luminosity. Synthetic intensity brightness maps were then derived from the scaled SED at the same units and wavelengths as the PACS data.  We then derived the synthetic intensity by integrating the flux in concentric apertures around the star and evaluating the difference between the fluxes in two successive apertures. The synthetic profiles were then combined and fitted to the observed intensity profiles and a best-fit model was selected for each star. In most cases, more than one detached shell was needed to fit the intensity profile (see Table~\ref{t:results}).
Our preliminary modelling result is shown in Fig.~\ref{profiles}. This could be improved by changing the number of detached shells and/or a modification of the density
law.
Additionally, we assumed that the terminal velocity of the gas is equal to the dust expansion velocity. For AQ And, a value of 15\,km~s$^{-1}$ was assumed. This assumption may be unrealistic as different shells for a star may expand at different velocities, but since the mass-loss rate scales linearly with the dust expansion velocity, we  can still obtain a rough estimate of the mass-loss rate. To compute the mass-loss rate, we used the relation from Groenewegen et al. (1998) applied to the case of a thin shell
$$\mathbf{\tau} _{\lambda} = \int_{r_{\rm d}}^{r_{out}} \! \pi a^2 Q_{\lambda} n(r) dr = 5.405\times10^{8} \frac{\dot{M}\ \Psi\ Q_{\lambda}/{a}}{r_{\rm d}\ R_*\ V_{\rm d}\ \rho_{\rm d}} \times (1-\frac{1}{Y_{\rm d}})\\,$$
where $\dot{M}$ is in M$_{\sun}$yr$^{-1}$, $V_{\rm d}$ is the dust velocity in km s$^{-1}$ , $R_*$ is the stellar radius in solar radii, $r_{\rm d}$ is the inner dust radius in stellar radii, $Q_{\lambda}$ is the absorption coefficient, $a$ is the dust grain radius in cm, $\Psi$ is the dust-to-gas mass ratio (we assumed 0.005), $\rho_{\rm d}$ is the grain density in g cm$^{-3}$, and $Y_{\rm d}$ is the shell's thickness in inner shell units.
The input for the models and the results are found in Table~\ref{t:results}. The true mass-loss may be lower, because the present wind velocity is probably lower than that assumed.

\begin{table*}
\scriptsize
\caption{Summary of the basic properties and fitting results.}
\label{t:results}
\centering
\begin{tabular}{l|cccccccc}
\hline
 & \multicolumn{3}{c|}{input parameters} & \multicolumn{5}{c}{best fit parameters} \\
\hline
 Object & $T_{\rm eff}$ [K]& $V_{gas}$ [Kms$^{-1}$]&  \multicolumn{1}{c|}{$D$ [pc]} & $\mathbf{\tau} _{\lambda}$  & $T_{\rm a}$  [K] & $T_{\rm d}$ [K] & $r_{\rm d}$ [\arcsec]& $Y_{\rm d}$ \\ 
\hline
 AQ\,And & 2700 & 15 & \multicolumn{1}{c|}{825}  & 0.003 &1200  & 27.9--27.5--29.1--30.5--32.4--40.4--44.8  & 51--48--41--35--30--14--10 &1.17--1.01--1.1--1.01--1.1--2--1.5   \\

 U\,Ant  & 2800  & 19 & \multicolumn{1}{c|}{260}  & 0.002&1000 & 39.3  & 40 & 1.3    \\
  
 TT\,Cyg & 2800 & 13.5 & \multicolumn{1}{c|}{510}  & 0.002&1000 & 28.3--30.6--34--36.8--45 & 33--26--19--14--7  & 1.16--1.13--1.11--1.11--2   \\
\hline
 & \multicolumn{6}{c}{models outputs} \\
 \hline
Object & $\dot{M}_a$ [M$_{\sun} {\rm yr}^{-1}$]& \multicolumn{3}{c}{$\dot{M}\ (\times$10$^{-7}$) [M$_{\sun} {\rm yr}^{-1}$]} & $L$ [ L$_{\sun}$]& ${M}\ (\times$10$^{-3}$) [M$_{\sun}$]    \\ 
\hline
 AQ\,And & 1$\times$10$^{-10}$ &  \multicolumn{3}{c}{74--152--9--57--3--0.6--0.3} &12\,000 & 20   \\

 U\,Ant  & 1$\times$10$^{-10}$ &  \multicolumn{3}{c}{7} &  8000   & 3.8$\times$10$^{-2}$   \\
  
 TT\,Cyg & 1.3$\times$10$^{-9}$ &  \multicolumn{3}{c}{2--3--0.2--0.1--0.09} & 2700  & 0.6   \\
 
\hline
\end{tabular}
\flushleft{\scriptsize{\textit{Note to Table~\ref{t:results}}: $T_{\rm eff}$ is the adopted effective temperature, $V_{gas}$ the terminal velocity derived from CO line measurements, $D$ the distance to the star, $\mathbf{\tau} _{\lambda}$ is the overall optical depth at 0.55$\mu$m, $T_{\rm a}$ the inner temperature of the attached shell, $T_{\rm d}$ the inner dust temperature of the detached shell(s), $r_{\rm d}$ the inner radius of the detached shell(s), $Y_{\rm d}$ the detached shell(s) thickness in inner shell radius units, $L$ the luminosity, $\dot{M}_a$ the actual mass-loss rate, $\dot{M}$ the mass-loss rate of the detached shell(s), and ${M}$ the total dust and gas mass-loss.}}
\vspace{-4mm}
\end{table*}

\section{Results} \label{res}

\textbf{AQ And}'s PACS images show a thin detached shell located at 52$\arcsec$ radius. In earlier surveys for CO radio line emission AQ~And was not detected. A new deep re-observation in CO (1$-$0) with the 20\,m telescope at Onsala Space Observatory resulted in a non detection (Olofsson, priv. comm.). This may be caused by photodissociation, keeping in mind the large physical extension of AQ~And's shell and its far distance. 
Using IRAS data, Young et al. (\cite{Young93a}, \cite{Young93b}) measured a source radius of about 3.3$\arcmin$ but this was only found at 60 $\mu$m.
An ISO-PHOT map at 60~$\mu$m detected extended emission comparable in size to the ring we see in our data when taking into account ISO-PHOT's lower resolution (online Fig.~\ref{f:AQAndISO}). There is no obvious extended emission at  3.3$\arcmin$ radius from the star neither in our maps nor in the ISO-PHOT maps. It is possible that the IRAS extended emission is caused by background contamination (CIRR2=3). Another possibility is that we filter out the very faint extended emission (see Sect. \ref{reduction}).
The inner part of the ring shows various intensity peaks (Fig.~\ref{profiles}). This points towards mass-loss variation over time. Seven models of detached shells with dust temperatures ranging from 28~K to 49~K were needed in order to reproduce the observed intensity profile. This corresponds to a mass-loss variation from 1$\times$10$^{-10}$M$_{\sun} {\rm yr}^{-1}$ for the present mass-loss to 152$\times$10$^{-7}$M$_{\sun} {\rm yr}^{-1}$ for the older mass-loss (some 19\,000 years ago). From the separation between the intensity peaks, we estimate that the mass-loss varies between every 2000 and 5000~yr. 

\onlfig{3}{
\begin{figure}
\begin{center}
\resizebox{!}{5.5cm}{\includegraphics{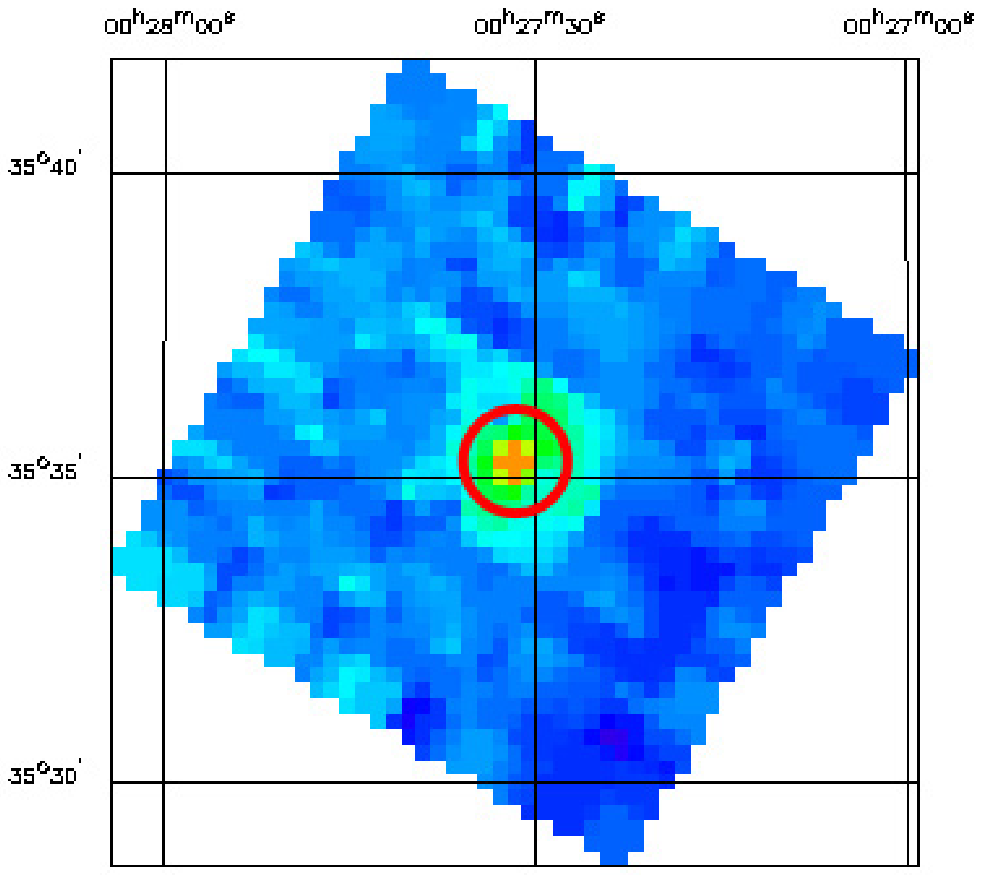}}
\end{center}
\caption{PHT22 ISO-PHOT observation of AQ And. 
A red circle denotes the shell (radius of 52\arcsec{}) observed with \textit{Herschel}-PACS discussed 
below. ISOs FWHM resolution is 35-40\arcsec{} at 60\,$\mu$m.}
\label{f:AQAndISO}
\end{figure}
}

\begin{figure}
\begin{center}
\vspace{-3mm}
\resizebox{!}{6cm}{\includegraphics[angle=0]{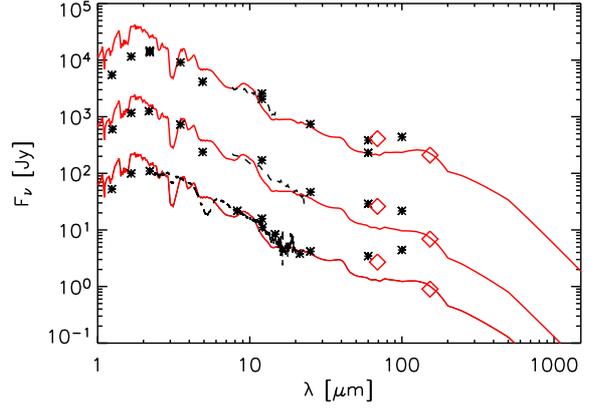}}
\end{center}
\vspace{-4mm}
\caption{SED fitting results. The dashed black line is the observed ISO short wavelength spectrum (SWS) and the long wavelength spectrum (LRS) (Sloan et al. 2003), the black symbols are literature photometry points, the red diamonds the PACS photometry, and the continuous red line is the model. From bottom to top: TT Cyg, 
U Ant, and AQ And, which was shifted by up by $10^2$\,Jy for clarity. }
\label{SED}
\end{figure}

\textbf{TT Cyg}'s intensity profile is similar to that of AQ\,And in that both indicate significant variation in the mass-loss rate. 
We note that the peak of the outer shell located about 33$\arcsec$ from the central star coincides remarkably well with the CO shell observed by Olofsson et al. (\cite{Olofs00}), though the CO shell is much thinner (2.5\arcsec) than the dust shell (see Fig.\ref{f:obs}).
Different scenarios have been proposed to explain the origin of the molecular shell. Olofsson et al. (\cite{Olofs90}) suggest a thermal pulse origin during which the star experienced a high mass-loss phase lasting only a few hundred years, 7000 years ago. A wind--wind interaction scenario is proposed by Sch\"oier et al. (\cite{Schoe05}) and Maercker et al. (2010). In this case, a rapidly moving shell is colliding with a slower moving wind sweeping up old material.
Another possible explanation mentioned by Wareing et al. (2006) is a wind--ISM interaction. For this to happen, the star's space velocity with respect to the ISM has to be high enough to induce a shock with the ISM. Bow shocks have been observed for AGB stars such as for R Hya (Ueta et al. 2006) and with PACS data for CW Leo (Ladjal et al. 2010). The shocked envelope may look circular if most of the motion is in the radial direction as is the case for TT\,Cyg's space velocity.
That there is no sign of the drift velocity between the dust and the gas envelopes, which one would expect to see if the detached shells were strictly of a thermal pulse origin, points more toward a wind--wind or a wind--ISM origin.
Five synthetic detached shells were needed to model the PACS intensity profiles (Fig.\ref{profiles}). The angular separation between the shells suggest that the mass-loss varied every $\sim$1500~yr.

\textbf{U Ant}'s intensity profile is very smooth with a detached shell located at 42$\arcsec$. This suggests that the star experienced a brief increase in mass-loss some 2800\,yr ago, after which the mass-loss rate dropped and has not varied much since. By imaging scattered light around U\,Ant, Gonz\`alez Delgado et al. (2001) proposed 4 shells at  $\sim$25\arcsec, 37\arcsec, 43\arcsec, and 46\arcsec\ (shell 1 to 4, respectively). Shell 4 was confirmed by studying scattered polarised light (Gonz\`alez Delgado et al. 2003). By assuming that the major fraction of polarised light comes from the dust and because of a difference in the polarisations of shells 3 and 4, they conclude that shell 4 consists mainly of dust, while shell 3 is gas dominated. Sch\"oier et al. (\cite{Schoe05}) modelled single-dish CO data and inferred that a thin CO shell  lies at the position of shell 3 ($\sim$43\arcsec).
Maercker et al. (\cite{Maerc10}) reobserved U\,Ant using scattered light and CO mapping and arrived at the same conclusion as Gonz\`alez Delgado et al. (2003) that shell 3 is at $\sim$43\arcsec\ and shell 4 at $\sim$50\arcsec.
The peak intensity of the dust shell seen using PACS coincides with shell 3 as seen by Gonz\`alez Delgado et al. (2003) and Maercker et al. (2010), a result that they attribute to gas.  We neither see nor resolve another shell at the position of shell 4.
Since the CO data confirm the existence of a detached molecular shell at $\sim$43\arcsec, we conclude that the dust shell and the gas shell are co-spatial. This is similar to what is seen for TT\,Cyg and again raises questions about the absence of gas/dust drift.
Concerning shell 4, one can suggest that it is not resolved in the PACS data, but if that were the case, the peak intensity that we see at $\sim$43\arcsec\ should have been clearly detected in polarised scattered light, but this is not the case.
The IRAS maps obtained 60 and 100~$\mu$m by Izumiura et al. (\cite{Izumi97}) indicate distant emission at about 3\arcmin. Only part of this extended emission could be recovered by special treatment of the maps (see online Fig.~\ref{UAntlow}). An improved data reduction may reveal all of the extended emission (see Sect. \ref{reduction}).
Using one detached shell model does not lead to a reliable fit to the intensity profile. Some of the emission seen within the outer shell is not accounted for. Varying the density distribution does not improve the fit and using more than one detached shell introduces a variation in the intensity that is not seen in the profile (Fig.\ref{profiles}). This discrepancy may be caused by the origin of the detached shell. If it did originate from a wind--wind or wind--ISM interaction, the envelope would have slowed down with time. This was not taken into account in our modelling.

\onlfig{5}{
\begin{figure}
\begin{center}
\resizebox{!}{4cm}{\includegraphics{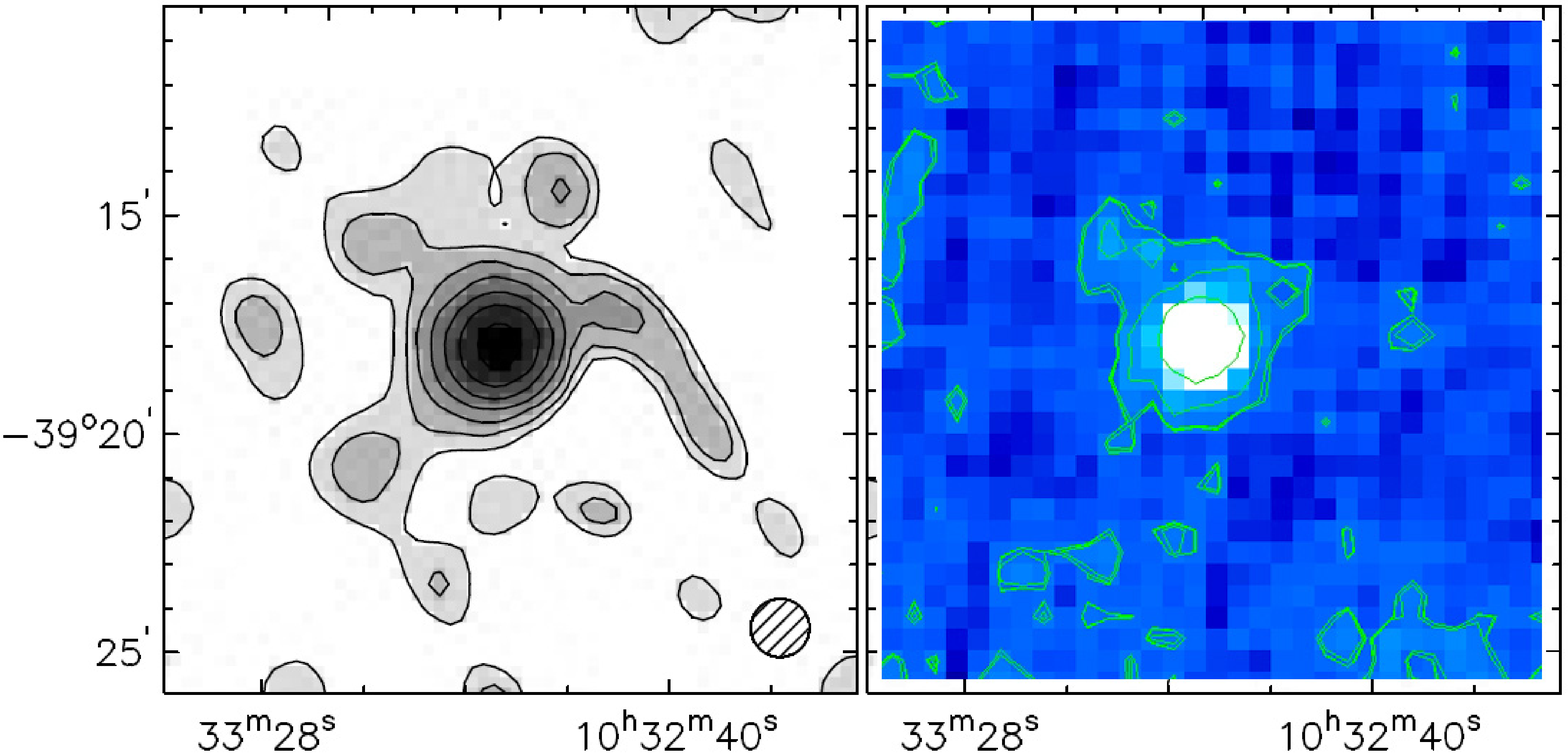}}
\end{center}
\caption{Extended field of U Ant: Left: HIRAS image at 100~$\mu$m (Izumiura et al. \cite{Izumi97}); right: \textit{Herschel}-PACS image at 70~$\mu$m rebinned to 30$\arcsec / $pixel.}
\label{UAntlow}
\end{figure}
}

\section{Conclusions}

All \textit{Herschel}-PACS maps detect circular rings of emission around the three central stars. These detached shells suggest that the stars underwent an increase in mass-loss for a short period of time. 
In all cases, we have found that varying dust emission continues to the inner parts of the detached shells. This can be seen from the  intensity profiles in Fig.~\ref{profiles} and the models. The corresponding timescales are of the order of thousands of years, whereas the age of the outer detached material is some 1000 or even 10\,000 years. 
The correlation with the molecular observations is striking in some cases even down to fine details, e.g., the lower shell density of TT~Cyg to the north. 
More detailed modelling combining all information about the objects, additional observations, and comparison with future \textit{Herschel} observations of similar objects will definitely provide a clearer understanding of the time evolution of the mass-loss process.

\begin{acknowledgements}
PACS has been developed by a consortium of institutes led by MPE
(Germany) and including UVIE (Austria); KUL, CSL, IMEC (Belgium); CEA, OAMP (France); MPIA (Germany); IFSI, OAP/AOT, OAA/CAISMI, LENS, SISSA (Italy); IAC (Spain). This development has been supported by the funding agencies BMVIT (Austria), ESA-PRODEX (Belgium), CEA/CNES (France), DLR (Germany), ASI (Italy), and CICYT/MCYT (Spain).
FK acknowledges funding by the Austrian Science Fund FWF under project number P18939-N16, RO under project number I163-N16. DL, MG, JB, WDM, KE, RH, CJ, RB, BV 
acknowledge support from the Belgian Federal Science Policy Office via the PRODEX Programme of ESA. The authors would like to thank the anonymous referee for constructive comments which led to a significant improvement of the paper.

\end{acknowledgements}

\end{document}